\documentclass[aps,prb,longbibliography,twocolumn]{revtex4-2}

\usepackage{graphicx}
\usepackage{dcolumn}
\usepackage{textcomp}
\usepackage[T1]{fontenc}
\usepackage{times}
\usepackage[colorlinks = true, linkcolor = blue, urlcolor  = blue, citecolor = blue, anchorcolor = blue]{hyperref}
\usepackage[english]{babel}
\usepackage{multirow}
\usepackage{dcolumn}
\usepackage{titlesec}
\titlespacing\section{0pt}{18pt plus 0pt minus 0pt}{6pt plus 0pt minus 0pt}
\titlespacing\subsection{0pt}{12pt plus 0pt minus 0pt}{6pt plus 0pt minus 0pt}
\begin{document}

\title{Quasiparticle band structure, spontaneous polarization, and spin-splitting in noncentrosymmetric few-layer and bulk $\gamma$-GeSe}
\author{Han-gyu Kim}
\author{Hyoung Joon Choi}
\email{h.j.choi@yonsei.ac.kr}
\address{Department of Physics, Yonsei University, Seoul 03722, Korea}
%\date{\today}

\begin{abstract}
Group-IV monochalcogenides have attracted much attention due to their potential of ferroelectric and multiferroic properties. 
Recently, centrosymmetric $\gamma$-phase GeSe in a double-layer honeycomb lattice has been theoretically predicted, 
but the synthesized $\gamma$-phase GeSe showed a noncentrosymmetric atomic structure, 
leading to the possibility of
%opening a chance for 
ferroelectricity and spin-splitting.
Here, we study the quasiparticle band structures,
spontaneous polarization, and spin-splitting in noncentrosymmetric $\gamma$-GeSe using density functional theory and GW calculations. 
Our results show that noncentrosymmetric few-layer and bulk $\gamma$-GeSe have
semiconducting band structures with indirect band gaps,
which depend almost linearly on the reciprocal of the number of layers.
Spontaneous polarization occurs due to a small charge transfer between the layers, which 
increases with compressive strain, and 
ferroelectric switching can be achieved by an interlayer translation with a small energy barrier.
Spin-splitting is found to be more significant at the highest valence band than at the lowest conduction band.
Our results provide insights into the fundamental electronic properties of a layered ferroelectric semiconductor
applicable to devices with ferroelectric/nonferroelectric junctions.
\end{abstract}

\maketitle

\section{Introduction}

Ferroelectric materials are of great interest because of their potential for various applications, such as nonvolatile memories 
\cite{Scott1989,Jiang2011,Naber2010}, 
field-effect transistors \cite{Mathews1997,Naber2005}, 
solar cells, sensors, and photonics devices \cite{Prateek2016,Wen2013,Catalan2009,Scott2007}. 
After the advent of semiconducting ferroelectric materials, ferroelectric thin films emerged as an important component in electronic applications \cite{Guan2020}. Recently, many two-dimensional materials with intrinsic ferroelectricity have been predicted by theoretical studies, such as 1T-MoS$_2$, In$_2$Se$_3$, and group-IV monochalcogenides \cite{Guan2020}.

It is known experimentally that group-IV monochalcogenides MX (M = Ge, Sn, X = S, Se, Te) have various atomic structures such as $\alpha$-phase 
\cite{Rohr2017,Vaughn2010,Bao2019,Zhao2014}, 
$\beta$-phase \cite{Rohr2017}, distorted rhombohedral structure \cite{Sante2013}, and simple rocksalt structure \cite{Hsieh2012}.
Group-IV monochalcogenides have been predicted to be intrinsic ferroelectric materials which have atomic structures of $\alpha$-phase \cite{Zeng2016,Fei2016}, 
$\beta$-phase \cite{Guan2018}, and $\gamma$-phase with a single-layer honeycomb lattice (SLHL) \cite{Liu2019}.
They have outstanding ferroelectric and multiferroic performances suited for device applications.

Recently, a new atomic structure of group-IV monochalcogenide $\gamma$-MX having a double-layer honeycomb lattice (DLHL) has been predicted theoretically \cite{Luo2020}. 
The predicted monolayer and AB-stacked bulk $\gamma$-GeSe are centrosymmetric, so 
the intrinsic ferroelectricity does not appear.
More recently, $\gamma$-GeSe has been synthesized experimentally \cite{KKim2020}
and found to have a slight difference in the atomic structure when compared with the theoretical prediction.
In the synthesized $\gamma$-GeSe \cite{KKim2020}, which we call AB$'$-stacked $\gamma$-GeSe, the stacking sequence of the atomic layers breaks the inversion symmetry, resulting in a noncentrosymmetric atomic structure and thereby 
leading to the possibility of
%opening a chance for 
ferroelectricity and spin-splitting.

In our present work, we study the
atomic and electronic structures, spontaneous polarization, and spin-splitting in noncentrosymmetric AB$'$-stacked
$\gamma$-GeSe using density functional theory (DFT) and GW calculations, and compare them with those in 
centrosymmetric AB-stacked $\gamma$-GeSe. 
We consider from a monolayer to a tetralayer and the bulk geometry of $\gamma$-GeSe using DFT calculations. 
Using GW calculations, we consider the monolayer, bilayer, and bulk $\gamma$-GeSe. 
We investigate the relaxed atomic structures, quasiparticle band structures, quasiparticle density of states (DOS), spontaneous polarization, strain dependence of spontaneous polarization, and spin-splitting in quasiparticle band structures 
in noncentrosymmetric AB$'$-stacked $\gamma$-GeSe, and we compare these results with those in centrosymmetric AB-stacked $\gamma$-GeSe.

\section{Computational methods}

We performed DFT calculations using the QUANTUM ESPRESSO code \cite{Giannozzi2009} with the Perdew-Burke-Ernzerhof-type generalized gradient approximation \cite{Perdew1996} for the exchange-correlation energy functional. We used a kinetic-energy cut-off of 100 Ry for the plane wave and norm-conserving pseudopotentials. 
For self-consistent calculations, we used a 12$\times$12$\times$3 $k$-point sampling for bulk and a 12$\times$12 $k$-point sampling in the two-dimensional Brillouin zone (BZ) of few-layer $\gamma$-GeSe.
The atomic structures of bulk and few-layer $\gamma$-GeSe were relaxed by minimizing the total energy of systems with the DFT-D2 scheme \cite{Grimme2006} which considers van der Waals interaction. For few-layer cases, we included a large enough vacuum region of 25~{\AA}.
We calculated the phonon dispersions of bulk $\gamma$-GeSe using density functional perturbation theory (DFPT) implemented in the QUANTUM ESPRESSO code.

\begin{figure} [t]
\includegraphics[scale=1]{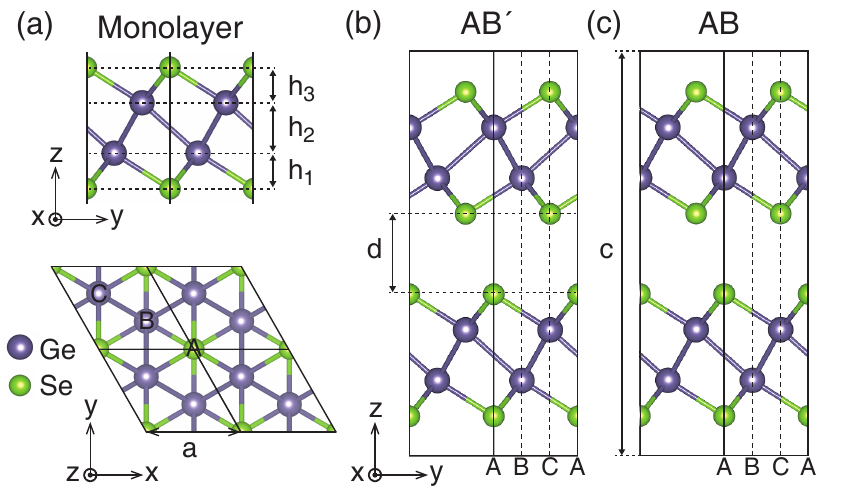}
\centering
\caption{\label{fig1gese}Atomic structures of (a) monolayer, (b) AB$'$-stacked, and (c) AB-stacked $\gamma$-GeSe. In (a), the top and side views are shown, where $a$ is the hexagonal lattice parameter and $h_1$, $h_2$, and $h_3$ are the distances between neighboring atomic layers inside a quadruple layer. In (b), $d$ is the distance between the quadruple layers. In (c), $c$ is the unit-cell length along the $c$ axis. In (b) and (c), A, B, and C indicate the sites in hexagonal lattice.}
\end{figure}

We calculated the quasiparticle band structures of bulk and few-layer $\gamma$-GeSe using the GW method implemented in the BerkeleyGW code \cite{Deslippe2012,Hybertsen1986,Rohlfing2000}. The GW method considers the electron-electron interaction using the self-energy, and describes the quasiparticle band structure accurately \cite{Hybertsen1986,Hedin1965,Hybertsen1985,Strinati1980,Strinati1982,Schilfgaarde2006,vanSetten2015,hgkim2021,hgkim2021a}.
It is known that the results of the GW method can depend on the
different levels of self-consistency and 
different methods to construct the initial starting 
electronic structure \cite{Hybertsen1986,Holm2004,Shishkin2007,Caruso2012,Atalla2013,vanSetten2015,Grumet2018,Foster2012}. 
The one-shot GW method calculates the self-energy only once from, 
and thereby depending on, the initial starting electronic 
structure, while the fully self-consistent GW method calculates 
the self-energy self-consistently, thus being independent of the 
initial starting electronic structure. 
In our present work, we used the one-shot GW method that 
calculates the self-energy only once from DFT band energies 
and wavefunctions, which is known to produce quasiparticle band 
structures in excellent agreement with experimental 
results \cite{Deslippe2012,Schilfgaarde2006,vanSetten2015,hgkim2021,hgkim2021a}.
For quasiparticle band structures including spin-orbit coupling, we considered spin-orbit coupling in calculating DFT bands, and then we shifted the DFT bands with the self-energy correction that is determined by the GW method without spin-orbit coupling. We used the Godby-Needs generalized plasmon pole model \cite{Godby1989, Oschlies1995} for the frequency-dependence of the inverse dielectric function. In our GW calculations, we used a kinetic-energy cut-off of 25 Ry for the dielectric matrix. We used a 4$\times$4$\times$1 uniform $q$-point sampling for the bulk system. For the case of few-layer $\gamma$-GeSe, we used 6$\times$6$\times$1 uniform $q$-point sampling with additional 10 $q$-points determined by the nonuniform neck subsampling (NNS) method \cite{Jornada2017}, which is equivalent to uniform 1143$\times$1143$\times$1 $q$ grids. We included 2000 bands for all cases with the static remainder method \cite{Deslippe2013}. 
With these parameters for GW calculations, the quasiparticle band energies are converged within 0.1~eV.

We used the Berry phase method \cite{smith1993} to calculate the spontaneous 
polarization of bulk $\gamma$-GeSe. 
For bilayer $\gamma$-GeSe, where in-plane spontaneous polarization is absent
due to the presence of multiple mirror planes 
perpendicular to the layer,
spontaneous polarization normal to the layer was calculated by 
integrating the charge density times the position vector.

\section{Results and discussion}

\begin{figure} [t]
\includegraphics[scale=1]{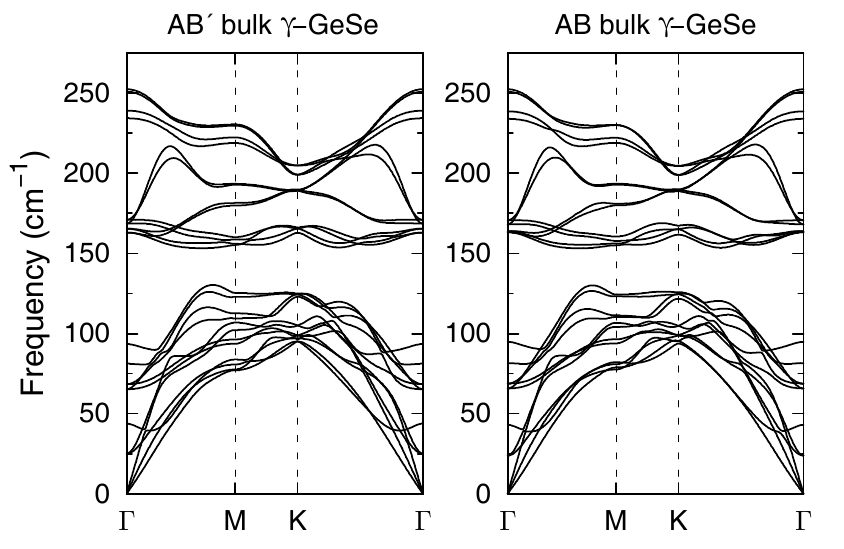}
\centering
\caption{\label{fig2gese}Phonon dispersions of (a) AB$'$- and (b) AB-stacked bulk $\gamma$-GeSe obtained from DFPT calculations.}
\end{figure}

Monolayer $\gamma$-GeSe is a quadruple layer of Se-Ge-Ge-Se in DLHL,
where four atomic layers Se, Ge, Ge, and Se are at A, B, C, and A sites of the hexagonal lattice, respectively, as shown in Fig.~\ref{fig1gese}(a). Previous theoretical work \cite{Luo2020} showed that this ABCA-sequence quadruple layer is stable and is energetically preferred to other stacking sequences such as ABAB, ABAC, and ABBA. 
For bulk $\gamma$-GeSe, one has to consider how quadruple layers are stacked, and it was reported that
AB stacking of quadruple layers is energetically preferred to AA stacking \cite{Luo2020}. 
Here, the AA stacking of quadruple layers in the bulk is the repetition of ABCA quadruple layers, and the AB stacking in the bulk is the alternation of ABCA quadruple and CABC quadruple layers, as shown in Fig.~\ref{fig1gese}(c).
For the bulk geometry containing two quadruple layers in the unit cell, we can also consider the alternation of the ABCA quadruple and CBAC quadruple layers, as shown in Fig.~\ref{fig1gese}(b).
This structure is consistent with recently synthesized $\gamma$-GeSe \cite{KKim2020}.
We refer to this atomic structure as AB$'$-stacked $\gamma$-GeSe. 

\begin{table*} [t]
\caption{\label{tab1gese}Relaxed structural parameters of $\gamma$-GeSe. The parameters $a$, $c$, $d$, $h_1$, $h_2$, and $h_3$ are defined in Fig.~\ref{fig1gese}. All values are in {\AA}. The first quadruple layer (QL) has the ABCA stacking of atomic layers. The second QL has the CABC stacking of atomic layers for the AB-stacked bilayer and bulk, the BCAB stacking of atomic layers for the BA-stacked bilayer, and the CBAC stacking for the AB$'$-stacked bilayer and bulk.}
\vspace*{5mm}
\renewcommand{\arraystretch}{1.12}
\begin{tabular*}{\textwidth}{@{\extracolsep{\fill}}lccccccccc}
\hline

 &\multirow{2}*{$a$}&\multirow{2}*{$c$}&\multirow{2}*{$d$}& \multicolumn{3}{c}{first QL} &  \multicolumn{3}{c}{second QL} \\
 &&   &   & $h_1$ & $h_2$ & $h_3$ & $h_1$ & $h_2$ & $h_3$ \\
\hline
Monolayer &	3.757 &	&	&1.356 	&1.950 	&1.356  & & &			\\
AB-bilayer	&3.759 &	& 3.090 	&1.355 	&1.931 	&1.356 	&1.356 	&1.931 	&1.355 \\
BA-bilayer	&3.763 &	& 3.080 	&1.352 	&1.928 	&1.352 	&1.352 	&1.928 	&1.352 \\
AB$'$-bilayer&	3.761 &&	3.076 &	1.354 	&1.931 	&1.353 	&1.355 	&1.934 	&1.354 \\
AB-bulk	&3.765 	&15.381 & 3.077	& 1.349 	&1.919 	&1.350 	&1.350 	&1.919 	&1.349 \\
AB$'$-bulk	&3.765 &	15.370 & 3.060 	&1.351 	&1.925 	&1.349 	&1.351 	&1.925 	&1.349 \\
\hline

\end{tabular*}
\end{table*}

For bilayer $\gamma$-GeSe, which consists of a total of eight atomic layers, 
we consider the atomic-layer sequences of ABCA+CABC, ABCA+BCAB, and ABCA+CBAC, and we call them as AB-, BA-, and AB$'$-bilayer, respectively.
Here, AB- and AB$'$-layers have the same atomic-layer sequence as their corresponding bulk while the atomic-layer sequence of ABCA+BCAB of the BA-bilayer, which is mapped to CABC+ABCA by in-plane translation,
corresponds to an exchange of two quadruple layers of the AB-bilayer. 
The AB- and BA-bilayers are not equivalent to each other because of the difference in the atomic registry between quadruple layers, so they have slightly different distances between quadruple layers.

In the case of the AB$'$-bilayer, the exchange of the two quadruple layers results in the atomic-layer sequence of CBAC+ABCA, which becomes ACBA+BCAB after in-plane translation and then it returns to ABCA+CBAC after 60$^\circ$ rotation around an axis normal to an atomic layer and passing through an A site. 
Thus, the exchange of the two quadruple layers of the AB$'$-bilayer does not yield a different geometry. 
Meanwhile, in the AB$'$-bilayer, in-plane translation of the CBAC quadruple layer to BACB produces ABCA+BACB and it becomes CABC+ACBA after overall in-plane translation, 
which is the overall reverse of the atom-layer sequence from ABCA+CBAC. 
Thus, the ABCA+CBAC and ABCA+BACB layers are mirror images of each other with respect to a plane parallel to an atomic layer, 
so the ABCA+CBAC and ABCA+BACB layers have the same distance between quadruple layers and their out-of-plane polarizations, if any, have the same size but opposite directions to each other.

Let N$_4$ be the number of quadruple layers in a two-dimensional slab of $\gamma$-GeSe. In general, AB- and BA-type stackings of quadruple layers are
centrosymmetric when N$_4$ is even. When N$_4$ is odd, inversion maps AB-type stacking to BA-type and {\em vice versa}. Meanwhile, AB$'$-type stacking of 
quadruple layers is always noncentrosymmetric, and in-plane translation of its B$'$ part from CBAC to BACB is equivalent to mirroring of the whole 
structure with respect to a plane parallel to an atomic layer followed by a translation parallel to the plane (when N$_4$ is even) or a rotation around an 
axis normal to the plane (when N$_4$ is odd).

In our present work, we study AB$'$-stacked $\gamma$-GeSe bulk and bilayer, and compare them with monolayers, AB-stacked $\gamma$-GeSe bulk, and AB- and BA-stacked $\gamma$-GeSe bilayers.
Our atomic structure relaxations with the DFT-D2 scheme
show that AB$'$-stacked bulk $\gamma$-GeSe is stable and it is energetically preferred over the AB-stacked bulk structure by 11~meV per unit cell. 
Here one unit cell contains eight atoms.
The stability of bulk $\gamma$-GeSe is further checked 
by calculating the phonon dispersions. 
The obtained phonon dispersions have zero frequency only for acoustic 
phonons at $\Gamma$, showing that both AB$'$- and AB-stacked bulk 
$\gamma$-GeSe are stable [Fig.~\ref{fig2gese}].
Phonon dispersions of AB$'$- and AB-stacked bulk $\gamma$-GeSe are slightly different
from each other because of the difference in their structural symmetries,
while those of the AB-stacked bulk [Fig.~\ref{fig2gese}(b)]
are consistent with the previous result \cite{Luo2020}. 
Our result that AB$'$-stacked bulk $\gamma$-GeSe is stable is 
consistent with the experimental report \cite{KKim2020}.

As for the symmetry of the atomic structure, the AB$'$-stacked bulk geometry belongs to space group 186 ($P{6}_{3}mc$), and AB-stacked bulk geometry belongs to space group 164 ($P\bar{3}m1$). The AB$'$-stacked bulk geometry is noncentrosymmetric and nonsymmorphic, while the AB-stacked bulk geometry is centrosymmetric and symmorphic. In the AB-stacked bilayer and bulk
geometries, the inversion center is at the middle of the interstitial region between the neighboring quadruple layers. The AB$'$-stacked bulk structure has the screw symmetry $S_{z}$ which is 180-degree rotation around the $c$-axis followed by half-unit-cell translation along the $c$-axis, and the glide mirror symmetry which includes the $c$-axis. The difference in the symmetry of AB- and AB$'$-stacked structures is due to the fact that the hexagonal stacking order in each quadruple layer is not changed by inversion but is changed by mirror or rotation preserving the $c$-axis. The monolayer $\gamma$-GeSe has inversion symmetry that belongs to the same point group as that of the AB-stacked bulk. Thus, the monolayer and AB-stacked $\gamma$-GeSe share some physical properties.

\begin{figure} [t]
\includegraphics[scale=1]{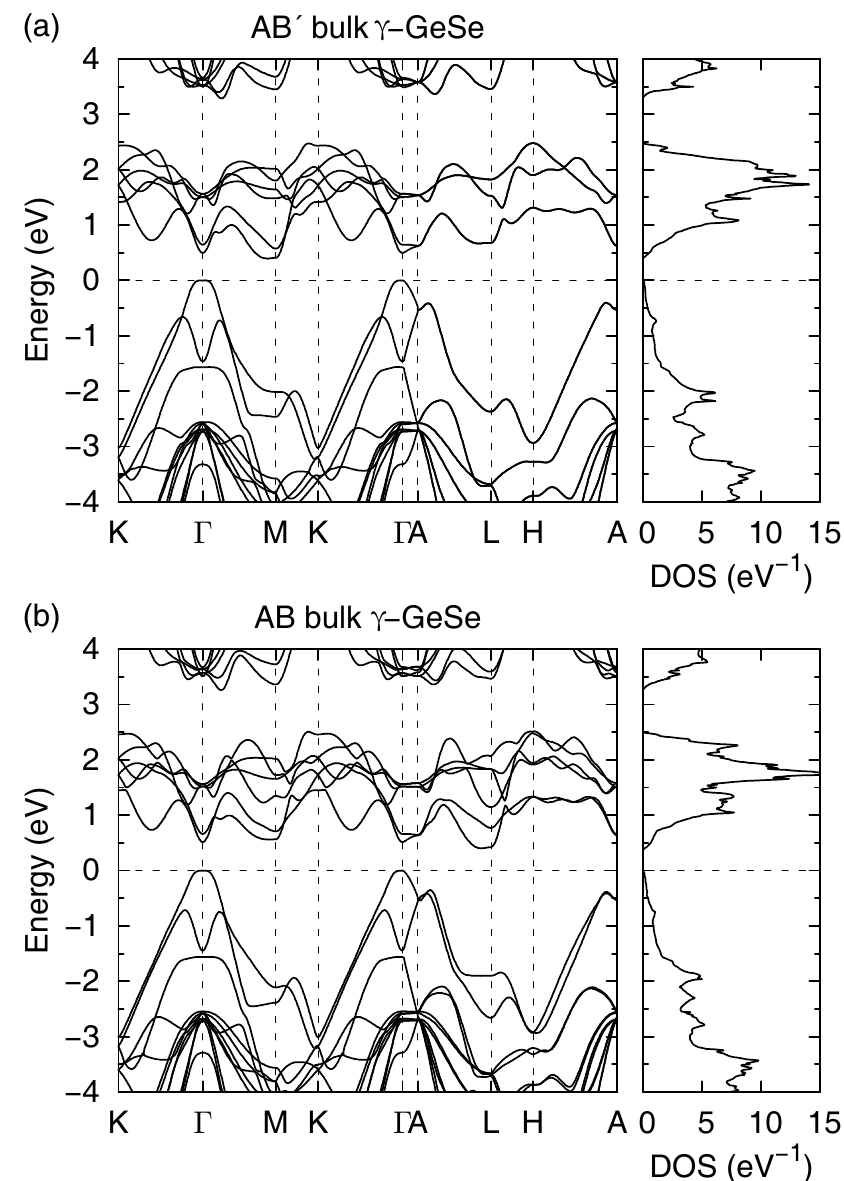}
\centering
\caption{\label{fig3gese}Quasiparticle band structures and the corresponding DOSs of (a) AB$'$-stacked and (b) AB-stacked bulk $\gamma$-GeSe obtained from GW calculations.}
\end{figure}

\begin{figure*} [t]
\includegraphics[scale=1.00]{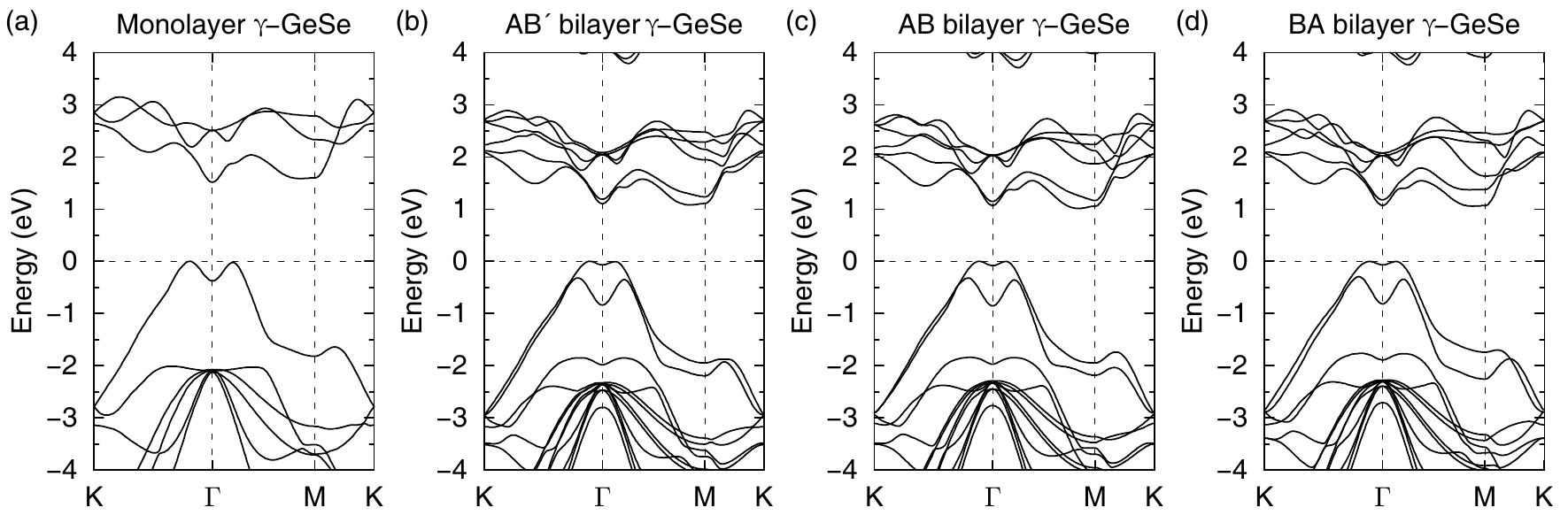}
\centering
\caption{\label{fig4gese}Quasiparticle band structures of (a) monolayer $\gamma$-GeSe, (b) AB$'$-, (c) AB-, and (d) BA-stacked bilayer $\gamma$-GeSe obtained from GW calculations.}
\end{figure*}

\begin{table} [t]
\centering
\caption{\label{tab2gese}Band gaps (in eV) of monolayer, bilayer, and bulk $\gamma$-GeSe.}
\vspace*{5mm}

\renewcommand{\arraystretch}{1.12}

\begin{tabular*}{0.48\textwidth}{@{\extracolsep{\fill}}lcccc}
\hline
	&\multicolumn{2}{c}{DFT band gaps}					&\multicolumn{2}{c}{GW band gaps}				\\
&	indirect	&direct		&indirect		&direct	 \\
\hline

Monolayer	&0.632 		&1.043 	&1.517 	&1.892  \\
AB-bilayer &	0.324 	&0.513 			& 1.014 &	1.152		 \\
BA-bilayer &	0.360 	&0.489 			& 1.059 &	1.140		 \\
AB$'$-bilayer & 	0.351 		&0.500 		& 1.081 & 1.172		\\
AB-bulk &	$-$0.038& 	0.109 	&0.404 	&0.517 	 \\
AB$'$-bulk &	$-$0.043&0.116 		&0.395 	&0.498 	 \\
\hline

\end{tabular*}

\end{table}

Table~\ref{tab1gese} shows the relaxed structural parameters of the monolayer, AB-stacked bilayer, 
BA-stacked bilayer,
AB$'$-stacked bilayer, AB-bulk, and AB$'$-bulk $\gamma$-GeSe. As the number of layers decreases, the lattice parameter $a$ decreases by 0.2\%. 
The distance $d$ between quadruple layers defined in the bilayer and bulk increases by 0.5\% as
the number of layers decreases. 
Distances $h_1$, $h_2$, and $h_3$ between atomic layers in each quadruple layer, as shown in Fig.~\ref{fig1gese}, also increase by 0.5\% as the number of layers decreases. In the monolayer geometry, $h_1$ and $h_3$ have the same value, with the inversion center at the middle of the quadruple layer. As for the different stacking, we find that $a$, $c$, $d$, $h_1$, $h_2$, and $h_3$ are different by less than 0.2\% between AB$'$- and AB-stacked geometries. Since the AB-stacked geometry has an inversion center at the middle of the interstitial region between neighboring quadruple layers, $h_1$($h_3$) of the first quadruple layer is equal to $h_3$($h_1$) of the second quadruple layer. Meanwhile, due to the screw symmetry $S_{z}$ in the AB$'$-stacked bulk geometry, $h_1$($h_3$) of the first quadruple layer is equal to $h_1$($h_3$) of the second quadruple layer. For the AB$'$-stacked bilayer, there is no rule for two quadruple layers to have an equal distance
between atomic layers, since there is no inversion symmetry or screw symmetry.

Fig.~\ref{fig3gese} shows the quasiparticle band structures and DOS of the AB$'$-stacked
and AB-stacked bulk $\gamma$-GeSe obtained by GW calculations without spin-orbit interaction. 
The AB$'$- and AB-stacked bulk are semiconducting in GW calculations as shown in Fig.~\ref{fig3gese}, although they are semimetallic in DFT calculations (Table~\ref{tab2gese}). 
In all structures under consideration in our present work, the VBM and CBM are at different $k$ points, that is, the band gap is indirect. We call this band gap as the indirect band gap hereafter. For comparison, we define the direct gap as the minimal energy difference between the conduction and valence bands at the same $k$ point. 
From our GW calculations, 
the indirect band gap of the AB$'$-stacked bulk $\gamma$-GeSe is 0.395~eV and its direct band gap is 0.498~eV, located at the $\Gamma$ point. 
In comparison, 
the indirect band gap of the AB-stacked bulk $\gamma$-GeSe is 0.404~eV and its direct band gap is 0.517~eV, located at the $\Gamma$ point. While the monolayer $\gamma$-GeSe has the highest valence-band dispersion resembling a camel's back \cite{Luo2020, Wang2014}, the two highest valence bands in bulk geometries are split greatly due to the interaction between quadruple layers so that the vicinity of the VBM becomes flat and the second highest valence band shows a deeper camel's back dispersion. 
Interestingly, we also note that the
valence bands show almost linear dispersion along the $K$-$\Gamma$ line in a large energy range from $-3$ to 0~eV. Since the band structures of the AB$'$- and AB-stacked bulk have low degeneracy around the band gap, their DOSs are small near the VBM and CBM as shown in Fig.~\ref{fig3gese}.

In quasiparticle band structures of the AB$'$- and AB- stacked bulk $\gamma$-GeSe, the positions of the VBM are almost the same, but the positions of the CBM are very different. 
In the quasiparticle band structure of the AB$'$-stacked bulk, 
the CBM is at a $k$-point in the $\Gamma$-$M$ line, but in that of the AB-stacked bulk, the CBM is at a $k$-point in the $A$-$L$ line at the BZ boundary. This difference is due to the difference of interactions between the quadruple layers in the AB$'$- and AB-stacked geometries. In the AB$'$-stacked bulk, the screw symmetry $S_{z}$ along the $c$-axis makes all bands doubly degenerate at the BZ boundary normal to the $c$-axis. One can show this by defining the operation $\Theta=TS_z$ and showing that $\Theta^2=-1$ at the BZ boundary \cite{Liang2016}, where $T$ is the time-reversal operator. This indicates that at the BZ boundary, the interaction between the quadruple layers results in nonbonding states.
Meanwhile, the interaction between the quadruple layers produces bonding and antibonding states in the $\Gamma$-$M$ line, with the CBM at a $k$-point in the line. 
In contrast, in the AB-stacked bulk $\gamma$-GeSe, larger splitting occurs
in the conduction bands at the BZ boundary, and as a result, the CBM is at a $k$-point in the $A$-$L$ line.

In addition to the AB$'$-stacked and AB-stacked bulk $\gamma$-GeSe, we also studied monolayer and bilayer $\gamma$-GeSe. Fig.~\ref{fig4gese} shows the quasiparticle band structures of mono- and bilayer $\gamma$-GeSe and a plot of band gaps {\em versus} the number of quadruple layers. 
Table~\ref{tab2gese} shows the indirect and direct band gaps from our DFT and GW calculations for monolayer, bilayer, and bulk $\gamma$-GeSe.
Fig.~\ref{fig4gese}(a) shows the quasiparticle band structure of monolayer $\gamma$-GeSe. The monolayer $\gamma$-GeSe has an indirect band gap of 1.517~eV, with the CBM located at the $\Gamma$ point and the VBM at a $k$-point in the $K$-$\Gamma$ line. 
The direct gap is 1.892~eV and is located near $\Gamma$.
Fig.~\ref{fig4gese}(b)-(d) show the quasiparticle band structures of AB$'$-stacked, AB-stacked, and BA-stacked bilayer $\gamma$-GeSe, respectively. 
The AB$'$-, AB-, and BA-stacked bilayers have indirect band gaps of 1.081, 1.014, and 1.059 eV, respectively.
The AB$'$-, AB-, and BA-stacked bilayers show similar band structures, in contrast to bulk $\gamma$-GeSe where the location of the CBM is different for AB$'$- and AB-stackings. In monolayer $\gamma$-GeSe, the highest valence band and the lowest conduction band have the almost same dispersion near the $\Gamma$ point, so the optical properties are expected to be excellent \cite{Luo2020}. 
In a multilayer geometry the camel's back-like dispersion becomes weak near the VBM, as observed in few-layer and bulk GaSe \cite{Cao2015,Li2014}.

\begin{figure} [t]
\includegraphics[scale=1.0]{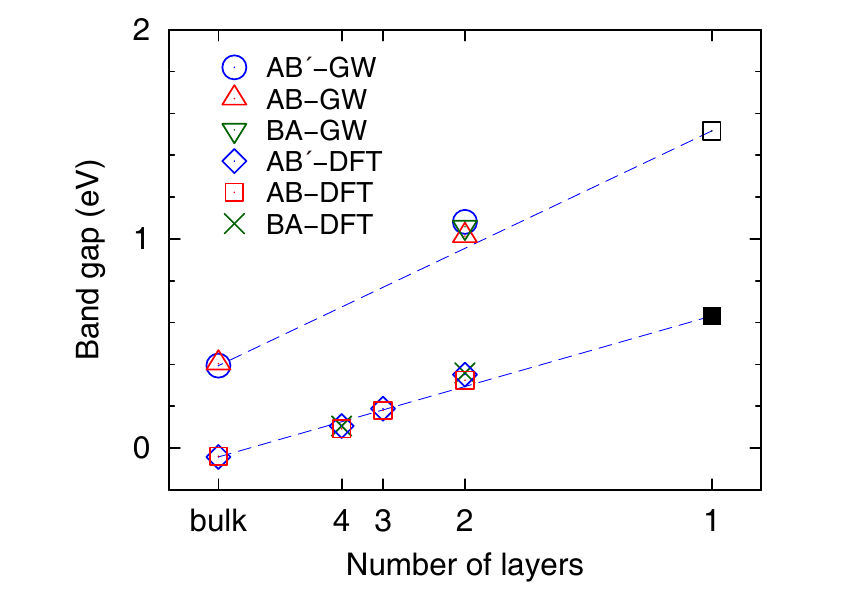}
\centering
\caption{\label{fig5gese} Indirect band gaps from DFT and GW calculations {\em versus} the number of quadruple layers. The black empty (filled) square dot shows a monolayer band gap from a GW (DFT) calculation. The horizontal axis is the reciprocal of the number of quadruple layers, and the upper (lower) dashed straight line connects the bulk and monolayer band gaps from GW (DFT) calculations.}
\end{figure} 

To find the thickness dependence of the band gap, we plotted indirect band gaps {\em versus} the reciprocal of the number of quadruple layers as shown in Fig.~\ref{fig5gese}, where we connected the band gaps of bulk and monolayer $\gamma$-GeSe with dashed straight lines. 
We note that in Fig.~\ref{fig5gese}, the DFT band gaps of bi-, tri-, and tetralayers are close to the straight line connecting the DFT band gaps of bulk and monolayer $\gamma$-GeSe. 
Fig.~\ref{fig5gese} also shows that the quasiparticle band gaps of bilayers are close to the straight line connecting those of bulk and monolayer $\gamma$-GeSe. Thus, using the straight line connecting the quasiparticle band gaps of bulk and monolayer $\gamma$-GeSe, we expect that quasiparticle band gaps of tri- and tetralayers are close to 0.77 and 0.68~eV, respectively.

Spontaneous polarization can occur in the AB$'$-stacked few-layer and bulk $\gamma$-GeSe because their atomic structures are noncentrosymmetric.
Group-IV monochalcogenides such as GeSe have already confirmed ferroelectricity in other various atomic structures \cite{Liu2019,Zeng2016,Fei2016,Guan2018}. 
We applied the Berry phase method \cite{smith1993} to the AB$'$-stacked bulk $\gamma$-GeSe and we simply integrated the charge density times the position vector along the out-of-plane direction for the AB$'$-stacked bilayer $\gamma$-GeSe. 
Spontaneous polarization does not exist in the in-plane direction due to the presence of multiple mirror planes containing the $c$-axis. 
Along the $c$-axis, spontaneous polarization of 
0.168~Debye per unit-cell exists for the AB$'$-stacked bulk and 
0.0173~Debye per unit-cell for the AB$'$-stacked bilayer.
These values correspond to
polarization per volume of 0.296~$\mu$C~cm$^{-2}$ for the AB$'$-stacked bulk $\gamma$-GeSe 
and polarization per area of 4.71$\times$10$^{-13}$~C~m$^{-1}$ for the AB$'$-stacked bilayer $\gamma$-GeSe.
These values are 
comparable to polarization values in
$\gamma$-GeSe with SLHL \cite{Liu2019}, 
traditional perovskite ferroelectric materials,
two-dimensional materials with a hexagonal buckling structure (SiGe, SiSn, GeSn, AlSb, GaP, InP, {\em etc}.) \cite{Sante2015},
and graphitic binary compound bilayers (BN, AlN, ZnO, MoS$_2$, GaSe, {\em etc}.) \cite{LeiLi2017}.
It is very interesting that $\gamma$-GeSe does or does not have the spontaneous polarization depending on the stacking method of quadruple layers although each constituent quadruple layer does not have any polarization if it is isolated.
This feature can bring in advantages for devices with ferroelectric/nonferroelectric junctions \cite{Huang2018,Chanthbouala2012,Garcia2009,Gruverman2009,Yau2017}.

In our stacking sequences in the AB$'$-stacked bulk and bilayer $\gamma$-GeSe as shown in Fig.~\ref{fig1gese}(b), polarizations are in the $-z$ direction. This direction of polarizations can be reversed by an interlayer translation as in graphitic binary compound bilayers \cite{LeiLi2017}. 
For example, a stack of ABCA+CBAC quadruple layers [Fig.~\ref{fig1gese}(b)] and a stack of ABCA+BACB quadruple layers are inverted to each other vertically, so polarizations of these structures are in opposite directions to each other. 
These stacking geometries are switchable by a translation of the second quadruple layer. By comparing the total energy of AB$'$-stacked $\gamma$-GeSe and intermediate states during the translation pathway, we obtained ferroelectric switching barriers of 71 and 19 meV per unit-cell for AB$'$-bulk and bilayer, respectively. 
These switching barriers are somewhat larger than that of a graphitic binary compound bilayer ($\sim$ 9~meV) \cite{LeiLi2017} due to the thicker quadruple-layer geometry, but they are much smaller than those of other ferroelectric materials, implying that ferroelectric device applications of the AB$'$-stacked $\gamma$-GeSe may have low energy consumption.

\begin{figure} [t]
\includegraphics[scale=1]{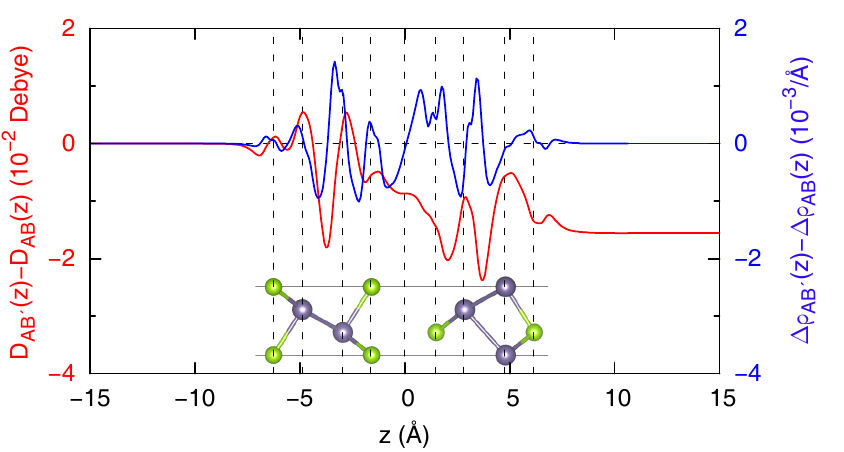}
\centering
\caption{\label{fig6gese}Difference in electric dipole moments and difference in electron densities of the AB$'$- and AB-stacked bilayer $\gamma$-GeSe. The red line shows the difference between D$_{\mathrm{AB}'}(z)$ of the AB$'$-stacked bilayer $\gamma$-GeSe and D$_\mathrm{AB}$($z$) of AB-stacked bilayer $\gamma$-GeSe. The blue line shows the difference between $\Delta\rho$$_{\mathrm{AB}'}$($z$) of the AB$'$-stacked bilayer $\gamma$-GeSe and $\Delta\rho$$_\mathrm{AB}$($z$) of AB-stacked bilayer $\gamma$-GeSe. D($z$) and $\Delta\rho$($z$) are defined in the text. The atomic model shows the AB$'$-stacked bilayer $\gamma$-GeSe in a slightly modified atomic structure where distances between atomic layers are made equal to those in the AB-stacked bilayer. The $z$ value is the position from the center of bilayer along the c-axis normal to layers.}
\end{figure}

To understand the origin of spontaneous polarization in the AB$'$-stacked bilayer $\gamma$-GeSe, we compared its electron distribution with
that of the AB-stacked $\gamma$-GeSe. 
We calculated the electric dipole moments of AB$'$- and AB-stacked bilayer $\gamma$-GeSe as a function of the out-of-plane coordinate $z$ from the center of the bilayer as
\begin{equation}
D({z}) = -e\int_{{z}_0}^{{z}} {z}' \Delta\rho({z}'){\mathrm d}{z}',
\end{equation}
where $e$ is the elementary charge ($e>0$), $\Delta{\rho}({z}')$ is the integration over the $z={z}'$ plane of the electron number density minus the superposition of the electron number densities of neutral atoms, and 
the lower limit $z_0$ of the integral is set to be $-$15~{\AA} 
from the center of bilayer $\gamma$-GeSe. If $z$ is large enough to include all electron densities, $D$($z$) is the total dipole moment of the bilayer. 
By calculating $D$($z=0$) and $D$(${z}=15$~{\AA}), 
we found that each quadruple layer in a bilayer is polarized
due to the increase of the electron density in the interstitial region between quadruple layers when compared with each isolated quadruple layer. 
In the AB-stacked bilayer $\gamma$-GeSe, 
the first quadruple layer has a dipole moment of 0.0488~Debye per unit-cell in the $-z$ direction and 
the second layer has a dipole moment of 0.0488~Debye per unit-cell in the $+z$ direction.
As the two dipole moments are of the same size but in the opposite directions, the total dipole moment is zero for the AB-stacked bilayer.

Fig.~\ref{fig6gese} shows the difference of dipole moments of AB$'$- and AB-stacked bilayer $\gamma$-GeSe as a function of the out-of-plane coordinate $z$.
In order to consider only the difference in the atomic-layer sequences in comparing
$D$($z$) of the AB$'$- and AB-stacked bilayer $\gamma$-GeSe, 
we modified the atomic structure of the AB$'$-stacked bilayer $\gamma$-GeSe
slightly to have
the same structural parameters $d$, $h_1$, $h_2$, and $h_3$ as the AB-stacked bilayer $\gamma$-GeSe,
and we used this modified structure to obtain the results that are plotted in Fig.~\ref{fig6gese}. 
As shown in Fig.~\ref{fig6gese}, the difference in $D$($z$) has a negative value at 
$z=0$ and a further negative value at $z=15$~{\AA}.
Fig.~\ref{fig6gese} also shows
$\Delta\rho_{\mathrm{AB}'}(z)-\Delta\rho_{\mathrm{AB}}(z)$
to compare the electron number density between the AB$'$- and AB-stacked bilayer $\gamma$-GeSe. 
These results indicate that the dipole moment of the first quadruple layer, which is in the $-z$ direction, has 18\% larger size in the AB$'$-stacked bilayer than the AB-stacked one while that of the second quadruple layer, which is in the $+z$ direction, has 14\% smaller size in the AB$'$-stacked one than the AB-stacked one.
This yields the total dipole moment in the $-z$ direction in the AB$'$-stacked bilayer $\gamma$-GeSe.

\begin{figure} [t]
\includegraphics[scale=1]{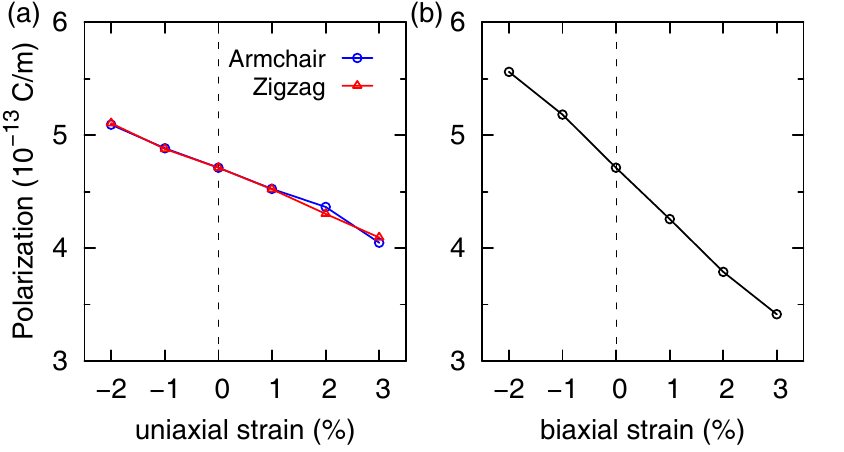}
\centering
\caption{\label{fig7gese}Spontaneous polarization of the AB$'$-stacked bilayer $\gamma$-GeSe as a function of (a) uniaxial strain along the armchair (blue circle) or zigzag direction (red triangle), and (b) biaxial strain. Sizes of the strain are represented with fractional changes.}
\end{figure}

To investigate any electron transfer between the quadruple layers, 
we integrated $\Delta\rho(z)$ along the out-of-plane direction from the left vacuum region ($z=-15$~{\AA})
to $z=0$ which is the middle of the interstitial region between quadruple layers. 
The integrated $\Delta\rho_{\mathrm{AB}}(z)$ for the first quadruple layer of the AB-stacked bilayer $\gamma$-GeSe is exactly zero due to the inversion symmetry centered at $z=0$, 
and the integrated $\Delta\rho_{\mathrm{AB}'}(z)$ for the first quadruple layer of the AB$'$-stacked bilayer $\gamma$-GeSe is $-$1.53$\times$10$^{-3}$ per unit-cell.
This value is almost the same with the integral of 
$\Delta\rho_{\mathrm{AB}'}(z)-\Delta\rho_{\mathrm{AB}}(z)$
from the topmost atomic layer ($z=-1.5$~{\AA}) of the first quadruple layer to $z=0$ along the out-of-plane direction.
These are consistent with the fact that the total dipole moment of the AB$'$-stacked bilayer is in the $-z$ direction,
showing that the polarization of the AB$'$-stacked bilayer is mainly due to 
the change in the electron distribution
in the interstitial region between quadruple layers, induced by broken inversion symmetry.

Mechanical flexibility is an advantage of two-dimensional materials, so strained atomic structures can be easily produced \cite{Peng2014,Guan2017}. We investigated the control of the polarization of $\gamma$-GeSe by introducing uniaxial or biaxial strain. As shown in Fig.~\ref{fig7gese}, we calculated the polarization of the AB$'$-stacked bilayer $\gamma$-GeSe as a function of uniaxial strain along armchair or zigzag direction and biaxial strain. We found that the polarization of the AB$'$-stacked bilayer $\gamma$-GeSe is enhanced by uniaxial and biaxial compressive strains while it is reduced by tensile strain. With the strain from $-$2\% to 3\%, 
the polarization per area varies from 5.09$\times$10$^{-13}$~C~m$^{-1}$ to 4.05$\times$10$^{-13}$~C~m$^{-1}$ for the uniaxial strain along the armchair direction, from 5.11$\times$10$^{-13}$~C~m$^{-1}$ to 4.09$\times$10$^{-13}$~C~m$^{-1}$ for the uniaxial strain along the zigzag direction, and from 5.56$\times$10$^{-13}$~C~m$^{-1}$ to 3.41$\times$10$^{-13}$~C~m$^{-1}$ for the biaxial strain. The effects of the two uniaxial strains are almost the same with a slight difference for the strain greater than 2\%, as shown in Fig.~\ref{fig7gese}(a). The effect of the biaxial strain is about twice the effect of the uniaxial strain.

\begin{figure} [t]
\includegraphics[scale=1]{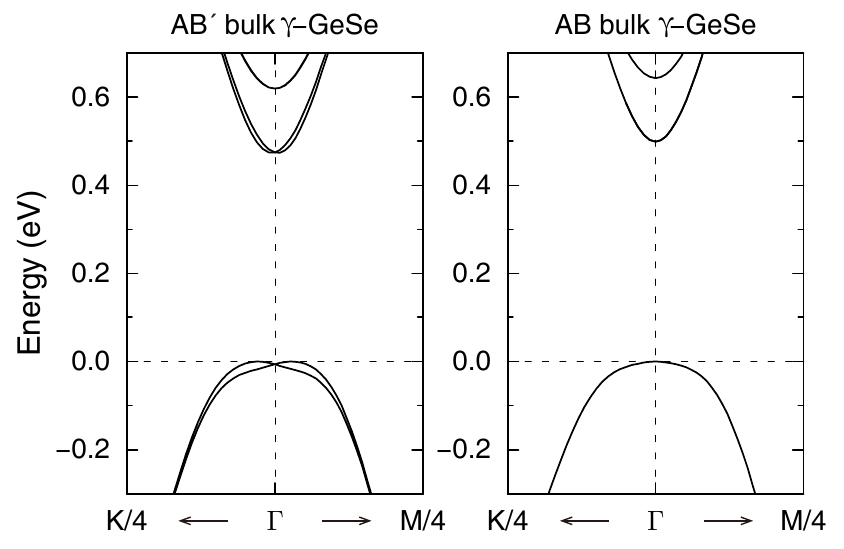}
\centering
\caption{\label{fig8gese}Spin-splitting in (a) AB$'$-stacked and (b) AB-stacked bulk $\gamma$-GeSe.}
\end{figure}

Finally, we analyze the effect of spin-orbit interaction in AB$'$-stacked bulk $\gamma$-GeSe.
Fig.~\ref{fig8gese} shows the quasiparticle band structure of AB$'$- and AB-stacked bulk $\gamma$-GeSe obtained by combining DFT calculations including spin-orbit coupling and GW calculations without spin-orbit coupling. It is well-known that materials with broken inversion symmetry have the Rashba- or Dresselhaus-type spin splitting near the time-reversal invariant momentum \cite{Acosta2019}. We found that the AB$'$-stacked bulk $\gamma$-GeSe has such a spin structure due to the absence of inversion symmetry. 
As shown in Fig.~\ref{fig8gese}(a), the spin-splitting is 20~meV at the highest valence band near the $\Gamma$ point, and the spin-splitting is much smaller at the lowest conduction band near the $\Gamma$ point. In comparison, the AB-stacked bulk $\gamma$-GeSe, which has inversion symmetry, has negligible spin-splitting at valence and conduction bands near the $\Gamma$ point [Fig.~\ref{fig8gese}(b)].

\section{Conclusion}
We studied the atomic and electronic structures, spontaneous polarization and its strain dependence, and spin-splitting in AB$'$-stacked $\gamma$-GeSe
using DFT and GW calculations,
and compared them with those in AB-stacked $\gamma$-GeSe.
From our DFT calculations, we found that the AB$'$-stacked $\gamma$-GeSe has lower total energy than the AB-stacked one.
The AB$'$-stacked bulk $\gamma$-GeSe is noncentrosymmetric and nonsymmorphic with screw symmetry and glide plane, while the AB-stacked bulk $\gamma$-GeSe is centrosymmetric and symmorphic.
We found that few-layer and bulk $\gamma$-GeSe have similar band gaps for the AB- and AB$'$-stacked geometries. 
However, for the case of bulk $\gamma$-GeSe, the locations of the CBM 
in the BZ are sensitive to stacking.
We also found that band gaps depend almost linearly on the reciprocal of the number of quadruple layers.
The screw symmetry $S_{z}$ along the $c$-axis in the AB$'$-stacked bulk 
makes bands degenerate at the BZ boundary, which are not degenerate in the AB-stacked bulk.
In the case of few-layer $\gamma$-GeSe, the differences of band dispersions and band gaps between AB- and AB$'$-stacked geometries were rather small. 
The AB$'$-stacked geometry has
spontaneous polarization along the out-of-plane direction 
regardless of the number of layers,
which originates from small electron transfer between quadruple layers.
The direction of polarization can be reversed by an interlayer translation with a small energy barrier.
When uniaxial or biaxial strain is applied to the
AB$'$-stacked bilayer $\gamma$-GeSe,
the spontaneous polarization decreases with compressive strain and increases with tensile strain.
The AB$'$-stacked bulk $\gamma$-GeSe shows a spin-splitting of 20 meV in the highest valence band near the $\Gamma$ point due to the absence of inversion symmetry, while the AB-stacked bulk $\gamma$-GeSe shows no spin-splitting.
Our findings provide basic information on a two-dimensional ferroelectric semiconductor which is applicable to devices with ferroelectric/nonferroelectric junctions.

\begin{acknowledgments}
This work is supported by the NRF of Korea (Grant No. 2020R1A2C3013673
and Grant No. 2017R1A5A1014862).
Computational resources have been provided by KISTI Supercomputing Center (Project No. KSC-2020-CRE-0335).
\end{acknowledgments}

%\bibliographystyle{apsrev4-2}

%\bibliography{manuscript_references}

\begin{mcitethebibliography}{62}
\providecommand*{\natexlab}[1]{#1}
\providecommand*{\mciteSetBstSublistMode}[1]{}
\providecommand*{\mciteSetBstMaxWidthForm}[2]{}
\providecommand*{\mciteBstWouldAddEndPuncttrue}
  {\def\EndOfBibitem{\unskip.}}
\providecommand*{\mciteBstWouldAddEndPunctfalse}
  {\let\EndOfBibitem\relax}
\providecommand*{\mciteSetBstMidEndSepPunct}[3]{}
\providecommand*{\mciteSetBstSublistLabelBeginEnd}[3]{}
\providecommand*{\EndOfBibitem}{}
\mciteSetBstSublistMode{f}
\mciteSetBstMaxWidthForm{subitem}
{(\emph{\alph{mcitesubitemcount}})}
\mciteSetBstSublistLabelBeginEnd{\mcitemaxwidthsubitemform\space}
{\relax}{\relax}

\makeatletter
\renewcommand\@biblabel[1]{#1}
\makeatother

\bibitem[Scott and Paz~de Araujo(1989)]{Scott1989}
J.~F. Scott and C.~A. Paz~de Araujo, {Ferroelectric Memories}, \emph{Science},
  1989, \textbf{246}, 1400--1405\relax
\mciteBstWouldAddEndPuncttrue
\mciteSetBstMidEndSepPunct{\mcitedefaultmidpunct}
{\mcitedefaultendpunct}{\mcitedefaultseppunct}\relax
\EndOfBibitem
\bibitem[Jiang \emph{et~al.}(2011)Jiang, Wang, Jin, Liu, Scott, Hwang, Tang,
  Lu, and Yang]{Jiang2011}
A.~Q. Jiang, C.~Wang, K.~J. Jin, X.~B. Liu, J.~F. Scott, C.~S. Hwang, T.~A.
  Tang, H.~B. Lu and G.~Z. Yang, {A Resistive Memory in Semiconducting
  BiFeO$_3$ Thin-Film Capacitors}, \emph{Adv. Mater.}, 2011, \textbf{23},
  1277--1281\relax
\mciteBstWouldAddEndPuncttrue
\mciteSetBstMidEndSepPunct{\mcitedefaultmidpunct}
{\mcitedefaultendpunct}{\mcitedefaultseppunct}\relax
\EndOfBibitem
\bibitem[Naber \emph{et~al.}(2010)Naber, Asadi, Blom, de~Leeuw, and
  de~Boer]{Naber2010}
R.~C.~G. Naber, K.~Asadi, P.~W.~M. Blom, D.~M. de~Leeuw and B.~de~Boer,
  {Organic Nonvolatile Memory Devices Based on Ferroelectricity}, \emph{Adv.
  Mater.}, 2010, \textbf{22}, 933--945\relax
\mciteBstWouldAddEndPuncttrue
\mciteSetBstMidEndSepPunct{\mcitedefaultmidpunct}
{\mcitedefaultendpunct}{\mcitedefaultseppunct}\relax
\EndOfBibitem
\bibitem[Mathews \emph{et~al.}(1997)Mathews, Ramesh, Venkatesan, and
  Benedetto]{Mathews1997}
S.~Mathews, R.~Ramesh, T.~Venkatesan and J.~Benedetto, {Ferroelectric Field
  Effect Transistor Based on Epitaxial Perovskite Heterostructures},
  \emph{Science}, 1997, \textbf{276}, 238--240\relax
\mciteBstWouldAddEndPuncttrue
\mciteSetBstMidEndSepPunct{\mcitedefaultmidpunct}
{\mcitedefaultendpunct}{\mcitedefaultseppunct}\relax
\EndOfBibitem
\bibitem[Naber \emph{et~al.}(2005)Naber, Tanase, Blom, Gelinck, Marsman,
  Touwslager, Setayesh, and de~Leeuw]{Naber2005}
R.~C.~G. Naber, C.~Tanase, P.~W.~M. Blom, G.~H. Gelinck, A.~W. Marsman, F.~J.
  Touwslager, S.~Setayesh and D.~M. de~Leeuw, {High-performance
  solution-processed polymer ferroelectric field-effect transistors},
  \emph{Nat. Mater.}, 2005, \textbf{4}, 243--248\relax
\mciteBstWouldAddEndPuncttrue
\mciteSetBstMidEndSepPunct{\mcitedefaultmidpunct}
{\mcitedefaultendpunct}{\mcitedefaultseppunct}\relax
\EndOfBibitem
\bibitem[{Prateek} \emph{et~al.}(2016){Prateek}, Thakur, and
  Gupta]{Prateek2016}
{Prateek}, V.~K. Thakur and R.~K. Gupta, {Recent Progress on Ferroelectric
  Polymer-Based Nanocomposites for High Energy Density Capacitors: Synthesis,
  Dielectric Properties, and Future Aspects}, \emph{Chem. Rev.}, 2016,
  \textbf{116}, 4260--4317\relax
\mciteBstWouldAddEndPuncttrue
\mciteSetBstMidEndSepPunct{\mcitedefaultmidpunct}
{\mcitedefaultendpunct}{\mcitedefaultseppunct}\relax
\EndOfBibitem
\bibitem[Wen \emph{et~al.}(2013)Wen, Li, Wu, Li, and Ming]{Wen2013}
Z.~Wen, C.~Li, D.~Wu, A.~Li and N.~Ming, Ferroelectric-field-effect-enhanced
  electroresistance in metal/ferroelectric/semiconductor tunnel junctions,
  \emph{Nat. Mater.}, 2013, \textbf{12}, 617--621\relax
\mciteBstWouldAddEndPuncttrue
\mciteSetBstMidEndSepPunct{\mcitedefaultmidpunct}
{\mcitedefaultendpunct}{\mcitedefaultseppunct}\relax
\EndOfBibitem
\bibitem[Catalan and Scott(2009)]{Catalan2009}
G.~Catalan and J.~F. Scott, {Physics and Applications of Bismuth Ferrite},
  \emph{Adv. Mater.}, 2009, \textbf{21}, 2463--2485\relax
\mciteBstWouldAddEndPuncttrue
\mciteSetBstMidEndSepPunct{\mcitedefaultmidpunct}
{\mcitedefaultendpunct}{\mcitedefaultseppunct}\relax
\EndOfBibitem
\bibitem[Scott(2007)]{Scott2007}
J.~F. Scott, {Applications of Modern Ferroelectrics}, \emph{Science}, 2007,
  \textbf{315}, 954--959\relax
\mciteBstWouldAddEndPuncttrue
\mciteSetBstMidEndSepPunct{\mcitedefaultmidpunct}
{\mcitedefaultendpunct}{\mcitedefaultseppunct}\relax
\EndOfBibitem
\bibitem[Guan \emph{et~al.}(2020)Guan, Hu, Shen, Xiang, Zhong, Chu, and
  Duan]{Guan2020}
Z.~Guan, H.~Hu, X.~Shen, P.~Xiang, N.~Zhong, J.~Chu and C.~Duan, {Recent
  Progress in Two-Dimensional Ferroelectric Materials}, \emph{Adv. Electron.
  Mater.}, 2020, \textbf{6}, 1900818\relax
\mciteBstWouldAddEndPuncttrue
\mciteSetBstMidEndSepPunct{\mcitedefaultmidpunct}
{\mcitedefaultendpunct}{\mcitedefaultseppunct}\relax
\EndOfBibitem
\bibitem[von Rohr \emph{et~al.}(2017)von Rohr, Ji, Cevallos, Gao, Ong, and
  Cava]{Rohr2017}
F.~O. von Rohr, H.~Ji, F.~A. Cevallos, T.~Gao, N.~P. Ong and R.~J. Cava,
  {High-Pressure Synthesis and Characterization of $\beta$-GeSe
  Six-Membered-Ring Semiconductor in an Uncommon Boat Conformation}, \emph{J.
  Am. Chem. Soc.}, 2017, \textbf{139}, 2771--2777\relax
\mciteBstWouldAddEndPuncttrue
\mciteSetBstMidEndSepPunct{\mcitedefaultmidpunct}
{\mcitedefaultendpunct}{\mcitedefaultseppunct}\relax
\EndOfBibitem
\bibitem[Vaughn \emph{et~al.}(2010)Vaughn, Patel, Hickner, and
  Schaak]{Vaughn2010}
D.~D. Vaughn, R.~J. Patel, M.~A. Hickner and R.~E. Schaak, {Single-Crystal
  Colloidal Nanosheets of GeS and GeSe}, \emph{J. Am. Chem. Soc.}, 2010,
  \textbf{132}, 15170--15172\relax
\mciteBstWouldAddEndPuncttrue
\mciteSetBstMidEndSepPunct{\mcitedefaultmidpunct}
{\mcitedefaultendpunct}{\mcitedefaultseppunct}\relax
\EndOfBibitem
\bibitem[Bao \emph{et~al.}(2019)Bao, Song, Liu, Chen, Zhu, Abdelwahab, Su, Fu,
  Chi, Yu, Liu, Zhao, Xu, Yang, and Loh]{Bao2019}
Y.~Bao, P.~Song, Y.~Liu, Z.~Chen, M.~Zhu, I.~Abdelwahab, J.~Su, W.~Fu, X.~Chi,
  W.~Yu, W.~Liu, X.~Zhao, Q.-H. Xu, M.~Yang and K.~P. Loh, {Gate-Tunable
  In-Plane Ferroelectricity in Few-Layer SnS}, \emph{Nano Lett.}, 2019,
  \textbf{19}, 5109--5117\relax
\mciteBstWouldAddEndPuncttrue
\mciteSetBstMidEndSepPunct{\mcitedefaultmidpunct}
{\mcitedefaultendpunct}{\mcitedefaultseppunct}\relax
\EndOfBibitem
\bibitem[Zhao \emph{et~al.}(2014)Zhao, Lo, Zhang, Sun, Tan, Uher, Wolverton,
  Dravid, and Kanatzidis]{Zhao2014}
L.-D. Zhao, S.-H. Lo, Y.~Zhang, H.~Sun, G.~Tan, C.~Uher, C.~Wolverton, V.~P.
  Dravid and M.~G. Kanatzidis, {Ultralow thermal conductivity and high
  thermoelectric figure of merit in SnSe crystals}, \emph{Nature}, 2014,
  \textbf{508}, 373--377\relax
\mciteBstWouldAddEndPuncttrue
\mciteSetBstMidEndSepPunct{\mcitedefaultmidpunct}
{\mcitedefaultendpunct}{\mcitedefaultseppunct}\relax
\EndOfBibitem
\bibitem[Di~Sante \emph{et~al.}(2013)Di~Sante, Barone, Bertacco, and
  Picozzi]{Sante2013}
D.~Di~Sante, P.~Barone, R.~Bertacco and S.~Picozzi, {Electric Control of the
  Giant Rashba Effect in Bulk GeTe}, \emph{Adv. Mater.}, 2013, \textbf{25},
  509--513\relax
\mciteBstWouldAddEndPuncttrue
\mciteSetBstMidEndSepPunct{\mcitedefaultmidpunct}
{\mcitedefaultendpunct}{\mcitedefaultseppunct}\relax
\EndOfBibitem
\bibitem[Hsieh \emph{et~al.}(2012)Hsieh, Lin, Liu, Duan, Bansil, and
  Fu]{Hsieh2012}
T.~H. Hsieh, H.~Lin, J.~Liu, W.~Duan, A.~Bansil and L.~Fu, {Topological
  crystalline insulators in the SnTe material class}, \emph{Nat. Commun.},
  2012, \textbf{3}, 982\relax
\mciteBstWouldAddEndPuncttrue
\mciteSetBstMidEndSepPunct{\mcitedefaultmidpunct}
{\mcitedefaultendpunct}{\mcitedefaultseppunct}\relax
\EndOfBibitem
\bibitem[Wu and Zeng(2016)]{Zeng2016}
M.~Wu and X.~C. Zeng, {Intrinsic Ferroelasticity and/or Multiferroicity in
  Two-Dimensional Phosphorene and Phosphorene Analogues}, \emph{Nano Lett.},
  2016, \textbf{16}, 3236--3241\relax
\mciteBstWouldAddEndPuncttrue
\mciteSetBstMidEndSepPunct{\mcitedefaultmidpunct}
{\mcitedefaultendpunct}{\mcitedefaultseppunct}\relax
\EndOfBibitem
\bibitem[Fei \emph{et~al.}(2016)Fei, Kang, and Yang]{Fei2016}
R.~Fei, W.~Kang and L.~Yang, {Ferroelectricity and Phase Transitions in
  Monolayer Group-IV Monochalcogenides}, \emph{Phys. Rev. Lett.}, 2016,
  \textbf{117}, 097601\relax
\mciteBstWouldAddEndPuncttrue
\mciteSetBstMidEndSepPunct{\mcitedefaultmidpunct}
{\mcitedefaultendpunct}{\mcitedefaultseppunct}\relax
\EndOfBibitem
\bibitem[Guan \emph{et~al.}(2018)Guan, Liu, Lu, Yao, and Yang]{Guan2018}
S.~Guan, C.~Liu, Y.~Lu, Y.~Yao and S.~A. Yang, {Tunable ferroelectricity and
  anisotropic electric transport in monolayer $\ensuremath{\beta}$-GeSe},
  \emph{Phys. Rev. B}, 2018, \textbf{97}, 144104\relax
\mciteBstWouldAddEndPuncttrue
\mciteSetBstMidEndSepPunct{\mcitedefaultmidpunct}
{\mcitedefaultendpunct}{\mcitedefaultseppunct}\relax
\EndOfBibitem
\bibitem[Liu \emph{et~al.}(2019)Liu, Guan, Yin, Wan, Wang, and Zhang]{Liu2019}
C.~Liu, S.~Guan, H.~Yin, W.~Wan, Y.~Wang and Y.~Zhang, {$\gamma$-GeSe: A
  two-dimensional ferroelectric material with doping-induced ferromagnetism},
  \emph{Appl. Phys. Lett.}, 2019, \textbf{115}, 252904\relax
\mciteBstWouldAddEndPuncttrue
\mciteSetBstMidEndSepPunct{\mcitedefaultmidpunct}
{\mcitedefaultendpunct}{\mcitedefaultseppunct}\relax
\EndOfBibitem
\bibitem[Luo \emph{et~al.}(2020)Luo, Duan, Yakobson, and Zou]{Luo2020}
N.~Luo, W.~Duan, B.~I. Yakobson and X.~Zou, {Excitons and Electron-Hole Liquid
  State in 2D $\gamma$-Phase Group-IV Monochalcogenides}, \emph{Adv. Funct.
  Mater.}, 2020, \textbf{30}, 2000533\relax
\mciteBstWouldAddEndPuncttrue
\mciteSetBstMidEndSepPunct{\mcitedefaultmidpunct}
{\mcitedefaultendpunct}{\mcitedefaultseppunct}\relax
\EndOfBibitem
\bibitem[Lee \emph{et~al.}(2021)Lee, Jung, Kim, Lee, Park, Jang, Yoon, Ghosh,
  Kim, Na, Kim, Choi, Cheong, and Kim]{KKim2020}
S.~Lee, J.-E. Jung, H.-g. Kim, Y.~Lee, J.~M. Park, J.~Jang, S.~H. Yoon,
  A.~Ghosh, M.~Kim, W.~Na, J.~H. Kim, H.~J. Choi, H.~Cheong and K.~Kim,
  {{$\gamma$-GeSe: A New Hexagonal Polymorph from Group IV-VI
  Monochalcogenide}}, \emph{Nano Lett.}, 2021, \textbf{21}, 4305\relax
\mciteBstWouldAddEndPuncttrue
\mciteSetBstMidEndSepPunct{\mcitedefaultmidpunct}
{\mcitedefaultendpunct}{\mcitedefaultseppunct}\relax
\EndOfBibitem
\bibitem[Giannozzi \emph{et~al.}(2009)Giannozzi, Baroni, Bonini, Calandra, Car,
  Cavazzoni, Ceresoli, Chiarotti, Cococcioni, Dabo, Corso, de~Gironcoli,
  Fabris, Fratesi, Gebauer, Gerstmann, Gougoussis, Kokalj, Lazzeri,
  Martin-Samos, Marzari, Mauri, Mazzarello, Paolini, Pasquarello, Paulatto,
  Sbraccia, Scandolo, Sclauzero, Seitsonen, Smogunov, Umari, and
  Wentzcovitch]{Giannozzi2009}
P.~Giannozzi, S.~Baroni, N.~Bonini, M.~Calandra, R.~Car, C.~Cavazzoni,
  D.~Ceresoli, G.~L. Chiarotti, M.~Cococcioni, I.~Dabo, A.~D. Corso,
  S.~de~Gironcoli, S.~Fabris, G.~Fratesi, R.~Gebauer, U.~Gerstmann,
  C.~Gougoussis, A.~Kokalj, M.~Lazzeri, L.~Martin-Samos, N.~Marzari, F.~Mauri,
  R.~Mazzarello, S.~Paolini, A.~Pasquarello, L.~Paulatto, C.~Sbraccia,
  S.~Scandolo, G.~Sclauzero, A.~P. Seitsonen, A.~Smogunov, P.~Umari and R.~M.
  Wentzcovitch, {QUANTUM} {ESPRESSO}: a modular and open-source software
  project for quantum simulations of materials, \emph{J. Phys.: Condens.
  Matter}, 2009, \textbf{21}, 395502\relax
\mciteBstWouldAddEndPuncttrue
\mciteSetBstMidEndSepPunct{\mcitedefaultmidpunct}
{\mcitedefaultendpunct}{\mcitedefaultseppunct}\relax
\EndOfBibitem
\bibitem[Perdew \emph{et~al.}(1996)Perdew, Burke, and Ernzerhof]{Perdew1996}
J.~P. Perdew, K.~Burke and M.~Ernzerhof, {Generalized Gradient Approximation
  Made Simple}, \emph{Phys. Rev. Lett.}, 1996, \textbf{77}, 3865--3868\relax
\mciteBstWouldAddEndPuncttrue
\mciteSetBstMidEndSepPunct{\mcitedefaultmidpunct}
{\mcitedefaultendpunct}{\mcitedefaultseppunct}\relax
\EndOfBibitem
\bibitem[Grimme(2006)]{Grimme2006}
S.~Grimme, {Semiempirical GGA-type density functional constructed with a
  long-range dispersion correction}, \emph{J. Comp. Chem.}, 2006, \textbf{27},
  1787--1799\relax
\mciteBstWouldAddEndPuncttrue
\mciteSetBstMidEndSepPunct{\mcitedefaultmidpunct}
{\mcitedefaultendpunct}{\mcitedefaultseppunct}\relax
\EndOfBibitem
\bibitem[Deslippe \emph{et~al.}(2012)Deslippe, Samsonidze, Strubbe, Jain,
  Cohen, and Louie]{Deslippe2012}
J.~Deslippe, G.~Samsonidze, D.~A. Strubbe, M.~Jain, M.~L. Cohen and S.~G.
  Louie, {BerkeleyGW: A massively parallel computer package for the calculation
  of the quasiparticle and optical properties of materials and nanostructures},
  \emph{Comput. Phys. Commun.}, 2012, \textbf{183}, 1269 -- 1289\relax
\mciteBstWouldAddEndPuncttrue
\mciteSetBstMidEndSepPunct{\mcitedefaultmidpunct}
{\mcitedefaultendpunct}{\mcitedefaultseppunct}\relax
\EndOfBibitem
\bibitem[Hybertsen and Louie(1986)]{Hybertsen1986}
M.~S. Hybertsen and S.~G. Louie, {Electron correlation in semiconductors and
  insulators: Band gaps and quasiparticle energies}, \emph{Phys. Rev. B}, 1986,
  \textbf{34}, 5390--5413\relax
\mciteBstWouldAddEndPuncttrue
\mciteSetBstMidEndSepPunct{\mcitedefaultmidpunct}
{\mcitedefaultendpunct}{\mcitedefaultseppunct}\relax
\EndOfBibitem
\bibitem[Rohlfing and Louie(2000)]{Rohlfing2000}
M.~Rohlfing and S.~G. Louie, {Electron-hole excitations and optical spectra
  from first principles}, \emph{Phys. Rev. B}, 2000, \textbf{62},
  4927--4944\relax
\mciteBstWouldAddEndPuncttrue
\mciteSetBstMidEndSepPunct{\mcitedefaultmidpunct}
{\mcitedefaultendpunct}{\mcitedefaultseppunct}\relax
\EndOfBibitem
\bibitem[Hedin(1965)]{Hedin1965}
L.~Hedin, {New Method for Calculating the One-Particle Green's Function with
  Application to the Electron-Gas Problem}, \emph{Phys. Rev.}, 1965,
  \textbf{139}, A796--A823\relax
\mciteBstWouldAddEndPuncttrue
\mciteSetBstMidEndSepPunct{\mcitedefaultmidpunct}
{\mcitedefaultendpunct}{\mcitedefaultseppunct}\relax
\EndOfBibitem
\bibitem[Hybertsen and Louie(1985)]{Hybertsen1985}
M.~S. Hybertsen and S.~G. Louie, {First-Principles Theory of Quasiparticles:
  Calculation of Band Gaps in Semiconductors and Insulators}, \emph{Phys. Rev.
  Lett.}, 1985, \textbf{55}, 1418--1421\relax
\mciteBstWouldAddEndPuncttrue
\mciteSetBstMidEndSepPunct{\mcitedefaultmidpunct}
{\mcitedefaultendpunct}{\mcitedefaultseppunct}\relax
\EndOfBibitem
\bibitem[Strinati \emph{et~al.}(1980)Strinati, Mattausch, and
  Hanke]{Strinati1980}
G.~Strinati, H.~J. Mattausch and W.~Hanke, {Dynamical Correlation Effects on
  the Quasiparticle Bloch States of a Covalent Crystal}, \emph{Phys. Rev.
  Lett.}, 1980, \textbf{45}, 290--294\relax
\mciteBstWouldAddEndPuncttrue
\mciteSetBstMidEndSepPunct{\mcitedefaultmidpunct}
{\mcitedefaultendpunct}{\mcitedefaultseppunct}\relax
\EndOfBibitem
\bibitem[Strinati \emph{et~al.}(1982)Strinati, Mattausch, and
  Hanke]{Strinati1982}
G.~Strinati, H.~J. Mattausch and W.~Hanke, {Dynamical aspects of correlation
  corrections in a covalent crystal}, \emph{Phys. Rev. B}, 1982, \textbf{25},
  2867--2888\relax
\mciteBstWouldAddEndPuncttrue
\mciteSetBstMidEndSepPunct{\mcitedefaultmidpunct}
{\mcitedefaultendpunct}{\mcitedefaultseppunct}\relax
\EndOfBibitem
\bibitem[van Schilfgaarde \emph{et~al.}(2006)van Schilfgaarde, Kotani, and
  Faleev]{Schilfgaarde2006}
M.~van Schilfgaarde, T.~Kotani and S.~Faleev, {Quasiparticle Self-Consistent
  $GW$ Theory}, \emph{Phys. Rev. Lett.}, 2006, \textbf{96}, 226402\relax
\mciteBstWouldAddEndPuncttrue
\mciteSetBstMidEndSepPunct{\mcitedefaultmidpunct}
{\mcitedefaultendpunct}{\mcitedefaultseppunct}\relax
\EndOfBibitem
\bibitem[van Setten \emph{et~al.}(2015)van Setten, Caruso, Sharifzadeh, Ren,
  Scheffler, Liu, Lischner, Lin, Deslippe, Louie, Yang, Weigend, Neaton, Evers,
  and Rinke]{vanSetten2015}
M.~J. van Setten, F.~Caruso, S.~Sharifzadeh, X.~Ren, M.~Scheffler, F.~Liu,
  J.~Lischner, L.~Lin, J.~R. Deslippe, S.~G. Louie, C.~Yang, F.~Weigend, J.~B.
  Neaton, F.~Evers and P.~Rinke, $GW$100: Benchmarking $G_0W_0$ for Molecular
  Systems, \emph{J. Chem. Theory Comput.}, 2015, \textbf{11}, 5665--5687\relax
\mciteBstWouldAddEndPuncttrue
\mciteSetBstMidEndSepPunct{\mcitedefaultmidpunct}
{\mcitedefaultendpunct}{\mcitedefaultseppunct}\relax
\EndOfBibitem
\bibitem[Kim and Choi(2021)]{hgkim2021}
H.-g. Kim and H.~J. Choi, {Thickness dependence of work function, ionization
  energy, and electron affinity of Mo and W dichalcogenides from DFT and GW
  calculations}, \emph{Phys. Rev. B}, 2021, \textbf{103}, 085404\relax
\mciteBstWouldAddEndPuncttrue
\mciteSetBstMidEndSepPunct{\mcitedefaultmidpunct}
{\mcitedefaultendpunct}{\mcitedefaultseppunct}\relax
\EndOfBibitem
\bibitem[Kim and Choi(2021)]{hgkim2021a}
H.-g. Kim and H.~J. Choi, Quasiparticle band structures of bulk and few-layer
  ${\mathrm{PdSe}}_{2}$ from first-principles $GW$ calculations, \emph{Phys.
  Rev. B}, 2021, \textbf{103}, 165419\relax
\mciteBstWouldAddEndPuncttrue
\mciteSetBstMidEndSepPunct{\mcitedefaultmidpunct}
{\mcitedefaultendpunct}{\mcitedefaultseppunct}\relax
\EndOfBibitem
\bibitem[Holm and von Barth(2004)]{Holm2004}
B.~Holm and U.~von Barth, Cancellation Effects in the {GW} Approximation,
  \emph{Phys. Scr.}, 2004, \textbf{T109}, 135\relax
\mciteBstWouldAddEndPuncttrue
\mciteSetBstMidEndSepPunct{\mcitedefaultmidpunct}
{\mcitedefaultendpunct}{\mcitedefaultseppunct}\relax
\EndOfBibitem
\bibitem[Shishkin and Kresse(2007)]{Shishkin2007}
M.~Shishkin and G.~Kresse, Self-consistent $GW$ calculations for semiconductors
  and insulators, \emph{Phys. Rev. B}, 2007, \textbf{75}, 235102\relax
\mciteBstWouldAddEndPuncttrue
\mciteSetBstMidEndSepPunct{\mcitedefaultmidpunct}
{\mcitedefaultendpunct}{\mcitedefaultseppunct}\relax
\EndOfBibitem
\bibitem[Caruso \emph{et~al.}(2012)Caruso, Rinke, Ren, Scheffler, and
  Rubio]{Caruso2012}
F.~Caruso, P.~Rinke, X.~Ren, M.~Scheffler and A.~Rubio, Unified description of
  ground and excited states of finite systems: The self-consistent $GW$
  approach, \emph{Phys. Rev. B}, 2012, \textbf{86}, 081102\relax
\mciteBstWouldAddEndPuncttrue
\mciteSetBstMidEndSepPunct{\mcitedefaultmidpunct}
{\mcitedefaultendpunct}{\mcitedefaultseppunct}\relax
\EndOfBibitem
\bibitem[Atalla \emph{et~al.}(2013)Atalla, Yoon, Caruso, Rinke, and
  Scheffler]{Atalla2013}
V.~Atalla, M.~Yoon, F.~Caruso, P.~Rinke and M.~Scheffler, Hybrid density
  functional theory meets quasiparticle calculations: A consistent electronic
  structure approach, \emph{Phys. Rev. B}, 2013, \textbf{88}, 165122\relax
\mciteBstWouldAddEndPuncttrue
\mciteSetBstMidEndSepPunct{\mcitedefaultmidpunct}
{\mcitedefaultendpunct}{\mcitedefaultseppunct}\relax
\EndOfBibitem
\bibitem[Grumet \emph{et~al.}(2018)Grumet, Liu, Kaltak,
  Klime\ifmmode~\check{s}\else \v{s}\fi{}, and Kresse]{Grumet2018}
M.~Grumet, P.~Liu, M.~Kaltak, J.~c.~v. Klime\ifmmode~\check{s}\else \v{s}\fi{}
  and G.~Kresse, Beyond the quasiparticle approximation: Fully self-consistent
  $GW$ calculations, \emph{Phys. Rev. B}, 2018, \textbf{98}, 155143\relax
\mciteBstWouldAddEndPuncttrue
\mciteSetBstMidEndSepPunct{\mcitedefaultmidpunct}
{\mcitedefaultendpunct}{\mcitedefaultseppunct}\relax
\EndOfBibitem
\bibitem[Foster and Wong(2012)]{Foster2012}
M.~E. Foster and B.~M. Wong, Nonempirically Tuned Range-Separated DFT
  Accurately Predicts Both Fundamental and Excitation Gaps in DNA and RNA
  Nucleobases, \emph{J. Chem. Theory Comput.}, 2012, \textbf{8},
  2682--2687\relax
\mciteBstWouldAddEndPuncttrue
\mciteSetBstMidEndSepPunct{\mcitedefaultmidpunct}
{\mcitedefaultendpunct}{\mcitedefaultseppunct}\relax
\EndOfBibitem
\bibitem[Godby and Needs(1989)]{Godby1989}
R.~W. Godby and R.~J. Needs, {Metal-insulator transition in Kohn-Sham theory
  and quasiparticle theory}, \emph{Phys. Rev. Lett.}, 1989, \textbf{62},
  1169--1172\relax
\mciteBstWouldAddEndPuncttrue
\mciteSetBstMidEndSepPunct{\mcitedefaultmidpunct}
{\mcitedefaultendpunct}{\mcitedefaultseppunct}\relax
\EndOfBibitem
\bibitem[Oschlies \emph{et~al.}(1995)Oschlies, Godby, and Needs]{Oschlies1995}
A.~Oschlies, R.~W. Godby and R.~J. Needs, {GW self-energy calculations of
  carrier-induced band-gap narrowing in n-type silicon}, \emph{Phys. Rev. B},
  1995, \textbf{51}, 1527--1535\relax
\mciteBstWouldAddEndPuncttrue
\mciteSetBstMidEndSepPunct{\mcitedefaultmidpunct}
{\mcitedefaultendpunct}{\mcitedefaultseppunct}\relax
\EndOfBibitem
\bibitem[da~Jornada \emph{et~al.}(2017)da~Jornada, Qiu, and Louie]{Jornada2017}
F.~H. da~Jornada, D.~Y. Qiu and S.~G. Louie, {Nonuniform sampling schemes of
  the Brillouin zone for many-electron perturbation-theory calculations in
  reduced dimensionality}, \emph{Phys. Rev. B}, 2017, \textbf{95}, 035109\relax
\mciteBstWouldAddEndPuncttrue
\mciteSetBstMidEndSepPunct{\mcitedefaultmidpunct}
{\mcitedefaultendpunct}{\mcitedefaultseppunct}\relax
\EndOfBibitem
\bibitem[Deslippe \emph{et~al.}(2013)Deslippe, Samsonidze, Jain, Cohen, and
  Louie]{Deslippe2013}
J.~Deslippe, G.~Samsonidze, M.~Jain, M.~L. Cohen and S.~G. Louie, {Coulomb-hole
  summations and energies for $GW$ calculations with limited number of empty
  orbitals: A modified static remainder approach}, \emph{Phys. Rev. B}, 2013,
  \textbf{87}, 165124\relax
\mciteBstWouldAddEndPuncttrue
\mciteSetBstMidEndSepPunct{\mcitedefaultmidpunct}
{\mcitedefaultendpunct}{\mcitedefaultseppunct}\relax
\EndOfBibitem
\bibitem[King-Smith and Vanderbilt(1993)]{smith1993}
R.~D. King-Smith and D.~Vanderbilt, {Theory of polarization of crystalline
  solids}, \emph{Phys. Rev. B}, 1993, \textbf{47}, 1651--1654\relax
\mciteBstWouldAddEndPuncttrue
\mciteSetBstMidEndSepPunct{\mcitedefaultmidpunct}
{\mcitedefaultendpunct}{\mcitedefaultseppunct}\relax
\EndOfBibitem
\bibitem[Wang \emph{et~al.}(2014)Wang, Wang, Liu, and Peng]{Wang2014}
X.~G. Wang, L.~Wang, J.~Liu and L.~M. Peng, {Camel-back band-induced power
  factor enhancement of thermoelectric lead-tellurium from Boltzmann transport
  calculations}, \emph{Appl. Phys. Lett.}, 2014, \textbf{104}, 132106\relax
\mciteBstWouldAddEndPuncttrue
\mciteSetBstMidEndSepPunct{\mcitedefaultmidpunct}
{\mcitedefaultendpunct}{\mcitedefaultseppunct}\relax
\EndOfBibitem
\bibitem[Liang \emph{et~al.}(2016)Liang, Zhou, Yu, Wang, and Weng]{Liang2016}
Q.-F. Liang, J.~Zhou, R.~Yu, Z.~Wang and H.~Weng, {Node-surface and node-line
  fermions from nonsymmorphic lattice symmetries}, \emph{Phys. Rev. B}, 2016,
  \textbf{93}, 085427\relax
\mciteBstWouldAddEndPuncttrue
\mciteSetBstMidEndSepPunct{\mcitedefaultmidpunct}
{\mcitedefaultendpunct}{\mcitedefaultseppunct}\relax
\EndOfBibitem
\bibitem[Cao \emph{et~al.}(2015)Cao, Li, and Louie]{Cao2015}
T.~Cao, Z.~Li and S.~G. Louie, {Tunable Magnetism and Half-Metallicity in
  Hole-Doped Monolayer GaSe}, \emph{Phys. Rev. Lett.}, 2015, \textbf{114},
  236602\relax
\mciteBstWouldAddEndPuncttrue
\mciteSetBstMidEndSepPunct{\mcitedefaultmidpunct}
{\mcitedefaultendpunct}{\mcitedefaultseppunct}\relax
\EndOfBibitem
\bibitem[Li \emph{et~al.}(2014)Li, Lin, Puretzky, Idrobo, Ma, Chi, Yoon,
  Rouleau, Kravchenko, Geohegan, and Xiao]{Li2014}
X.~Li, M.-W. Lin, A.~A. Puretzky, J.~C. Idrobo, C.~Ma, M.~Chi, M.~Yoon, C.~M.
  Rouleau, I.~I. Kravchenko, D.~B. Geohegan and K.~Xiao, {Controlled Vapor
  Phase Growth of Single Crystalline, Two-Dimensional GaSe Crystals with High
  Photoresponse}, \emph{Sci. Rep.}, 2014, \textbf{4}, 5497\relax
\mciteBstWouldAddEndPuncttrue
\mciteSetBstMidEndSepPunct{\mcitedefaultmidpunct}
{\mcitedefaultendpunct}{\mcitedefaultseppunct}\relax
\EndOfBibitem
\bibitem[Di~Sante \emph{et~al.}(2015)Di~Sante, Stroppa, Barone, Whangbo, and
  Picozzi]{Sante2015}
D.~Di~Sante, A.~Stroppa, P.~Barone, M.-H. Whangbo and S.~Picozzi, {Emergence of
  ferroelectricity and spin-valley properties in two-dimensional honeycomb
  binary compounds}, \emph{Phys. Rev. B}, 2015, \textbf{91}, 161401\relax
\mciteBstWouldAddEndPuncttrue
\mciteSetBstMidEndSepPunct{\mcitedefaultmidpunct}
{\mcitedefaultendpunct}{\mcitedefaultseppunct}\relax
\EndOfBibitem
\bibitem[Li and Wu(2017)]{LeiLi2017}
L.~Li and M.~Wu, {Binary Compound Bilayer and Multilayer with Vertical
  Polarizations: Two-Dimensional Ferroelectrics, Multiferroics, and
  Nanogenerators}, \emph{ACS Nano}, 2017, \textbf{11}, 6382--6388\relax
\mciteBstWouldAddEndPuncttrue
\mciteSetBstMidEndSepPunct{\mcitedefaultmidpunct}
{\mcitedefaultendpunct}{\mcitedefaultseppunct}\relax
\EndOfBibitem
\bibitem[Huang \emph{et~al.}(2018)Huang, Yau, Yu, Qi, So, Dai, and
  Jiang]{Huang2018}
S.~Huang, H.-M. Yau, H.~Yu, L.~Qi, F.~So, J.-Y. Dai and X.~Jiang,
  {Flexoelectricity in a metal/ferroelectric/semiconductor heterostructure},
  \emph{AIP Advances}, 2018, \textbf{8}, 065321\relax
\mciteBstWouldAddEndPuncttrue
\mciteSetBstMidEndSepPunct{\mcitedefaultmidpunct}
{\mcitedefaultendpunct}{\mcitedefaultseppunct}\relax
\EndOfBibitem
\bibitem[Chanthbouala \emph{et~al.}(2012)Chanthbouala, Crassous, Garcia,
  Bouzehouane, Fusil, Moya, Allibe, Dlubak, Grollier, Xavier, Deranlot, Moshar,
  Proksch, Mathur, Bibes, and Barth{\'e}l{\'e}my]{Chanthbouala2012}
A.~Chanthbouala, A.~Crassous, V.~Garcia, K.~Bouzehouane, S.~Fusil, X.~Moya,
  J.~Allibe, B.~Dlubak, J.~Grollier, S.~Xavier, C.~Deranlot, A.~Moshar,
  R.~Proksch, N.~D. Mathur, M.~Bibes and A.~Barth{\'e}l{\'e}my, {Solid-state
  memories based on ferroelectric tunnel junctions}, \emph{Nat. Nanotech.},
  2012, \textbf{7}, 101--104\relax
\mciteBstWouldAddEndPuncttrue
\mciteSetBstMidEndSepPunct{\mcitedefaultmidpunct}
{\mcitedefaultendpunct}{\mcitedefaultseppunct}\relax
\EndOfBibitem
\bibitem[Garcia \emph{et~al.}(2009)Garcia, Fusil, Bouzehouane, Enouz-Vedrenne,
  Mathur, Barthlmy, and Bibes]{Garcia2009}
V.~Garcia, S.~Fusil, K.~Bouzehouane, S.~Enouz-Vedrenne, N.~D. Mathur,
  A.~Barthlmy and M.~Bibes, {Giant tunnel electroresistance for non-destructive
  readout of ferroelectric states}, \emph{Nature}, 2009, \textbf{460},
  81--84\relax
\mciteBstWouldAddEndPuncttrue
\mciteSetBstMidEndSepPunct{\mcitedefaultmidpunct}
{\mcitedefaultendpunct}{\mcitedefaultseppunct}\relax
\EndOfBibitem
\bibitem[Gruverman \emph{et~al.}(2009)Gruverman, Wu, Lu, Wang, Jang, Folkman,
  Zhuravlev, Felker, Rzchowski, Eom, and Tsymbal]{Gruverman2009}
A.~Gruverman, D.~Wu, H.~Lu, Y.~Wang, H.~W. Jang, C.~M. Folkman, M.~Y.
  Zhuravlev, D.~Felker, M.~Rzchowski, C.-B. Eom and E.~Y. Tsymbal, {Tunneling
  Electroresistance Effect in Ferroelectric Tunnel Junctions at the Nanoscale},
  \emph{Nano Lett.}, 2009, \textbf{9}, 3539--3543\relax
\mciteBstWouldAddEndPuncttrue
\mciteSetBstMidEndSepPunct{\mcitedefaultmidpunct}
{\mcitedefaultendpunct}{\mcitedefaultseppunct}\relax
\EndOfBibitem
\bibitem[Yau \emph{et~al.}(2017)Yau, Xi, Chen, Wen, Wu, and Dai]{Yau2017}
H.-M. Yau, Z.~Xi, X.~Chen, Z.~Wen, G.~Wu and J.-Y. Dai, {Dynamic strain-induced
  giant electroresistance and erasing effect in ultrathin ferroelectric
  tunnel-junction memory}, \emph{Phys. Rev. B}, 2017, \textbf{95}, 214304\relax
\mciteBstWouldAddEndPuncttrue
\mciteSetBstMidEndSepPunct{\mcitedefaultmidpunct}
{\mcitedefaultendpunct}{\mcitedefaultseppunct}\relax
\EndOfBibitem
\bibitem[Peng \emph{et~al.}(2014)Peng, Wei, and Copple]{Peng2014}
X.~Peng, Q.~Wei and A.~Copple, {Strain-engineered direct-indirect band gap
  transition and its mechanism in two-dimensional phosphorene}, \emph{Phys.
  Rev. B}, 2014, \textbf{90}, 085402\relax
\mciteBstWouldAddEndPuncttrue
\mciteSetBstMidEndSepPunct{\mcitedefaultmidpunct}
{\mcitedefaultendpunct}{\mcitedefaultseppunct}\relax
\EndOfBibitem
\bibitem[Guan \emph{et~al.}(2017)Guan, Yu, Liu, Liu, Dong, Lu, Yao, and
  Yang]{Guan2017}
S.~Guan, Z.-M. Yu, Y.~Liu, G.-B. Liu, L.~Dong, Y.~Lu, Y.~Yao and S.~A. Yang,
  {Artificial gravity field, astrophysical analogues, and topological phase
  transitions in strained topological semimetals}, \emph{npj Quantum Mater.},
  2017, \textbf{2}, 23\relax
\mciteBstWouldAddEndPuncttrue
\mciteSetBstMidEndSepPunct{\mcitedefaultmidpunct}
{\mcitedefaultendpunct}{\mcitedefaultseppunct}\relax
\EndOfBibitem
\bibitem[Mera~Acosta \emph{et~al.}(2019)Mera~Acosta, Fazzio, and
  Dalpian]{Acosta2019}
C.~Mera~Acosta, A.~Fazzio and G.~M. Dalpian, {Zeeman-type spin splitting in
  nonmagnetic three-dimensional compounds}, \emph{npj Quantum Mater.}, 2019,
  \textbf{4}, 41\relax
\mciteBstWouldAddEndPuncttrue
\mciteSetBstMidEndSepPunct{\mcitedefaultmidpunct}
{\mcitedefaultendpunct}{\mcitedefaultseppunct}\relax
\EndOfBibitem
\end{mcitethebibliography}

%\providecommand{\newblock}{}
%\begin{thebibliography}{10}
%\expandafter\ifx\csname url\endcsname\relax
%  \def\url#1{{\tt #1}}\fi
%\expandafter\ifx\csname urlprefix\endcsname\relax\def\urlprefix{URL }\fi
%\providecommand{\eprint}[2][]{\url{#2}}

\providecommand*{\mcitethebibliography}{\thebibliography}
\csname @ifundefined\endcsname{endmcitethebibliography}
{\let\endmcitethebibliography\endthebibliography}{}

%\clearpage
%\end{thebibliography}{10}

\end{document}